\documentclass[sigconf,10pt,twocolumn,nonacm]{acmart}
\acmSubmissionID{152}
\usepackage{algorithm}
\usepackage{algpseudocode}
\usepackage{placeins}
\begin{document}

\newcommand{\ignore}[1]{}
\newcommand{\enote}[1]{{\color{blue}\begin{quote}{\bf Erez's note:} #1\end{quote}}}
\newcommand{\tnote}[1]{{\color{magenta}\begin{quote}{\bf Tomer's note:} #1\end{quote}}}
\renewcommand{\algorithmiccomment}[1]{\textcolor{blue!45}{$//$ #1}}

\title{Skiplists with Foresight: Skipping Cache Misses}
%\title{Cache-Conscious Skiplists with Foresight}
%\title{Skipping Cache Misses in Skiplists}

%\title{Foresight: a Skiplist Optimization to Increase Throughput and Reduce Cache Misses}
\author{Tomer Cory}
\orcid{0009-0008-6977-0341}
\affiliation{%
  \institution{Technion}
  \city{}
  \country{Israel}
}
\email{tomer.cory@campus.technion.ac.il}

\author{Niv Sulimany}
\orcid{TODO}
\affiliation{%
  \institution{Technion}
  \city{}
  \country{Israel}
}
\email{sulimany.niv@campus.technion.ac.il}

\author{Erez Petrank}
\orcid{0000-0002-6353-956X}
\affiliation{%
  \institution{Technion}
  \city{}
  \country{Israel}
}
\email{erez@cs.technion.ac.il}

\renewcommand{\shortauthors}{Tomer Cory, Niv Sulimany and Erez Petrank}


\begin{abstract}
A skiplist is a fundamental data structure widely used in systems and applications for indexing data stores.  
In this work, we introduce {\em Foresight}, a cache-friendly skiplist optimization. 
Extending {\em Foresight} to concurrent settings introduces significant synchronization challenges that we identify and address.
{\em Foresight} is a surgical optimization, easy to integrate into a wide variety of skiplist designs. We apply it to one sequential and three concurrent skiplist designs and observe throughput improvements of up to 45\% in microbenchmarks.
When applied to a skiplist-based index in the DBx1000 in-memory database, {\em Foresight} yields end-to-end performance gains of up to 15\%.
\end{abstract}

\maketitle % should come after the abstract
\section{Introduction}

A skiplist, first introduced by Pugh~\cite{pugh1990skip}, is a probabilistic data structure often used as a simpler alternative to balanced search trees~\cite{AVL,B-trees,red-black}, supporting search, insert, and delete operations with expected complexity of $O(\log n)$. In addition to sets and maps, skiplists have also been used to implement priority queues~\cite{shavit2000skiplist, linden2013skiplist}.
Their relative ease of implementation, especially in the concurrent case, in multi-core environments where multiple threads may access the data structure concurrently, and their efficiency (especially for range queries) compared to balanced search trees have led to their wide adoption in a variety of applications, including many database engines and storage systems~\cite{leveldb, rocksdb, singlestore, hbase,  redis, memcached, kannan2018redesigning, zhang2019s3, xie2016pi} where they serve as in-memory key-value stores, as well as applications in areas such as blockchains~\cite{nikitin2017chainiac,8751340}, search~\cite{lucene}, streaming~\cite{wang2006peer}, and even synchronization primitives~\cite{murthy2016design}.

In this work, we present {\sl Foresight}, a skiplist optimization that can be applied to a variety of skiplist designs. {\sl Foresight} improves locality, thereby reducing data movement, which is a major performance bottleneck in modern architectures. The core idea is to store a copy of each node’s key in its predecessor node at every level. This allows a thread to determine whether to continue a traversal right (i.e., in the same level) or down (i.e., in a lower level) without first dereferencing the next node in the level  to check its key. Avoiding this extra visit to the next node significantly improves skiplist performance, because this extra visit often incurs a cache miss. This simple method can be easily incorporated into a sequential skiplist, and it provides a significant performance advantage, as shown in our evaluation. 
Storing keys in predecessor nodes introduces space overhead. This cost may be only partially noticeable  in practical key-value stores where the space held by values can be significantly 
larger than the space taken by the index.

Extending this method to a concurrent skiplist is non-trivial because the next pointer and the foreseen key of the successor node cannot be read atomically together. Concurrent modifications by other threads can cause the foreseen key to become inconsistent with the actual key of the referenced node, which may violate the correctness of skiplist operations. We present two techniques to overcome this challenge and evaluate them on multiple concurrent skiplist designs.
%\enote{You need to shortly talk also about the first method. As it stands now it doesn't flow well. Maybe something like: The first method employs standard synchronization techniques including... The other technique ...}
%\tnote{Added a short presentation below.}
The first technique utilizes optimistic access to retain {\sl Foresight}'s benefits, coupled with an efficient validation step to ensure correctness without compromising performance. The second leverages the atomicity of SIMD (Single Instruction, Multiple Data) double-word loads and stores on modern architectures. This unconventional utilization of SIMD instructions enables reading both the pointer and the foreseen key values atomically, eliminating the inconsistent foreseen key problem. To the best of our knowledge, this is the first reported use of SIMD instructions for synchronization in the literature. The evaluation demonstrates that this approach can provide superior performance compared to the alternative synchronization method. Notably, both synchronization techniques preserve the progress guarantees of the underlying skiplist.

We implemented and evaluated {\sl Foresight} using the Synchrobench benchmarking suite~\cite{gramoli2015more}, showing a significant throughput increase. We observe a clear correlation between this performance gain and a {\sl Foresight}-induced reduction in cache misses per skiplist operation across all cache levels. 
To evaluate real-world impact, we integrated {\sl Foresight} into skiplist-based indexes within the DBx1000 in-memory database~\cite{yu2014staring} and observed end-to-end performance improvements of up to 15\% across representative database workloads.

{\sl Foresight} is a simple and surgical optimization, making it applicable to many modern (and potentially future) skiplist designs, and also leading to a very lightweight integration: in our implementations, applying {\sl Foresight} required modifying only 81-130 lines of code, depending on the skiplist design.

{\textit{{Our Contributions:}}} %This paper makes the following contributions:
\noindent\begin{enumerate}
    \item Presenting the {\sl Foresight} optimization for skiplists.
    \item Extending the optimization to support concurrent skiplists.
    \item Leveraging the atomicity of SIMD instructions on modern architectures to simplify synchronization. 
    \item Implementing and evaluating {\sl Foresight} on a sequential skiplist and three concurrent skiplist designs from the Synchrobench benchmarking suite, and a skiplist index in the DBx1000 in-memory database.  
\end{enumerate}

% \textit{{Paper Organization:}} 
% Section~\ref{section:preliminary} describes our computational model and provides necessary background on the skiplist data structure. We define the {\sl Foresight} optimization in Section \ref{sec::foresight} and discuss its integration into skiplists, focusing on synchronization challenges.
% Our experimental results are presented in Section \ref{Evaluation}. Related work is discussed in Section \ref{Related Work}, and Section \ref{Conclusion} concludes this paper.

\section{Preliminaries}\label{section:preliminary}

Throughout this work, we consider a standard shared-memory setting in which a set of threads access shared memory using the atomic primitives \texttt{read}, \texttt{write}, and \texttt{compare\allowbreak-and-swap}. The \texttt{compare-and-swap} (CAS) operation conditionally updates a memory location's value if it matches an expected value. These atomic instructions are widely supported on modern processors. Moreover, many modern processors also provide a wide CAS instruction that conditionally updates two adjacent words atomically, provided both the length and alignment are 16 bytes.

A concurrent data structure's execution is said to be {\em linearizable} when each of its operations appears to be performed atomically at a certain point between its invocation and response, termed its {\em linearization point}, in accordance with the sequential specification of the data structure. A concurrent data structure is {\em linearizable} if all its possible executions meet this criterion~\cite{herlihy:linearizability}. All the concurrent skiplist designs mentioned in this paper are linearizable. 

A concurrent data structure is {\em lock-free} if whenever threads are performing data structure operations, some thread is guaranteed to complete its operation within a finite amount of computation steps. A concurrent data structure is {\em wait-free} if whenever a thread is performing a data structure operation, that thread is guaranteed to complete its operation within a finite amount of computation steps.

%A concurrent object is said to be {\em lock-free} if, at any point in time, at least one of the threads trying to access the object makes progress within a finite number of steps, even if other threads are paused or delayed. 
%Furthermore, a concurrent object framework is {\em wait-free} when any operation by any thread can be completed within a finite number of steps, without being influenced by the operational speed of other threads~\cite{herlihy1991wait}.

A {\em set} represents a collection of distinct keys. It supports the following operations: the \texttt{insert($k$)} function inserts the key $k$ unless it is already present, in which case it returns an error. The symmetric \texttt{delete($k$)} function removes $k$ if found; otherwise, it returns an error. The \texttt{contains($k$)} function verifies the presence of $k$ within the set. %!rewrite
A {\em dictionary}, also known as a {\em map}, represents a collection of distinct keys, each paired with a corresponding value. 
%Its operations resemble those of a set, with the difference of incorporating values. 
Its operations resemble those of a set, except that each key is paired with a value.
In this paper, we discuss skiplists, a data structure that is often used to implement sets and dictionaries. As items are indexed according to their keys in both cases, traversals and modifications are performed similarly. Thus, we do not make clear distinctions between the two.
%We will mainly discuss sets in this paper, yet the observations and assertions made here are also true for dictionaries. 

%\section{The Skiplist Data Structure}\label{sec:skiplist}
%Let us review the design of skiplists as the basis for this work. 

\subsection{The Skiplist Data Structure}
Skiplists consist of multiple levels of linked lists ordered by element keys. The bottom level contains all skiplist elements, while each higher level contains a subset of the elements from the level below. Each level begins with a sentinel node whose key is defined as $-\infty$, smaller than any element’s key.

Level $0$ of a skiplist is often called the data layer, while the upper levels are referred to collectively as the index layer. Multiple occurrences of the same element across levels form a tower. In most skiplist implementations, tower heights are chosen randomly during insertion, typically according to a geometric distribution. The probabilistic skiplists in Synchrobench use $G(\tfrac{1}{2})$, meaning that half of the towers have height $1$, a quarter have height $2$, and so on. Since heights are bounded, the expected tower height is at most $2$.
There are two approaches to constructing skiplist towers:

\textit{\textbf{Array-based towers}}. In this approach, each skiplist element is represented by a single node that maintains a \textit{next} array. This array stores the pointers to the element’s right neighbor at every level in which the node appears.
            \ignore{towers are constructed as a single node with a 
            \textit{next} array whose $i$-th cell holds the pointer to the next node in the $i$-th level list.}
            Examples of skiplists that follow this approach are those described in \cite{pugh1990skip, herlihy2007simple, fraser2004practical}. 
            
\textit{\textbf{Linked-list-based towers}}. In this approach, the skiplist is organized as a two-dimensional linked list. Each node corresponds to a particular element at a specific level and maintains pointers to both the node directly below and the node to its right.
    \ignore{and contains a \textit{next} field, pointing to the next node at that level, and a \textit{down} field, pointing to the node that represents the same element in the level below. Often, in this approach, different types of nodes are used for nodes in the data layer and index layer such that only the data layer nodes contain information like elements' keys and values. When this is the case, index layer nodes will also contain a pointer to the data node of their element.}
Examples of skiplists that follow this approach can be found in \cite{openjdk-skiplist-src, crain2013no}.

        \ignore{\tnote{Maybe I should avoid this discussion at this point, and just talk about towers as an abstract thing. I could come back to it when presenting the different skiplist implementations relevant for our work.}}
In array-based towers, a single contiguous node represents an element across multiple levels (in contrast to one node per level in linked-list-based towers). This design typically provides better locality, resulting in superior performance~\cite{lee2019low}.
On the other hand, linked-list-based towers are more versatile: they allow the use of different algorithms to manage the list in the data layer and the lists in the index layers~\cite{openjdk-skiplist-src}, and they facilitate dynamically changing an element’s height (i.e., the number of levels in which it appears)~\cite{crain2013no}.
While some of the skiplists discussed in this paper are array-based and some are linked-list-based, for simplicity, the pseudocode throughout this work assumes that the skiplist towers are array-based.
    We move to describing 
    skiplist operations in the sequential case, following Pugh's skiplist design~\cite{pugh1990skip}.

        \begin{figure*}[bhtp]
          \centering
            \includegraphics[width=\linewidth]{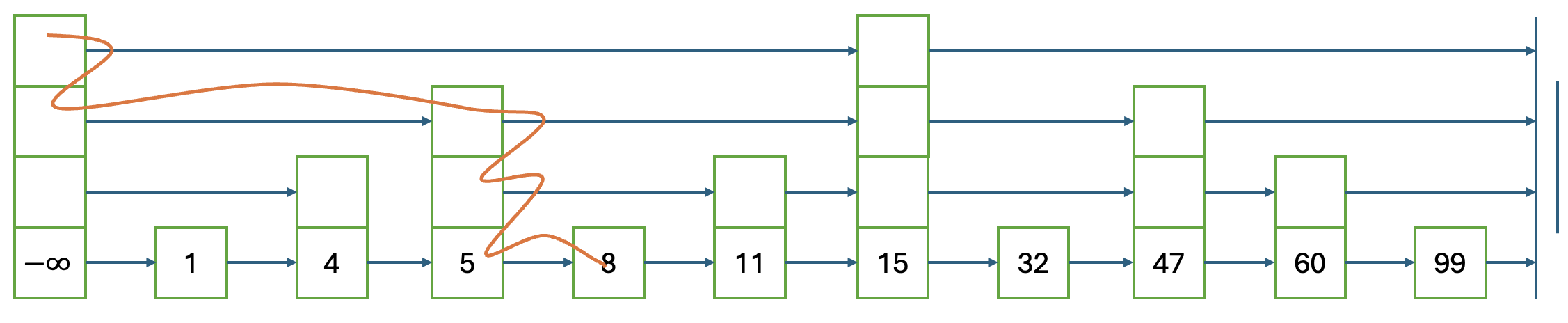}
          \caption{A skiplist example. The search path for elements with $5<key\leq 8$ is marked.} \label{skiplist figure}
          \Description{A skiplist example. The search path for elements with $8<key\leq 11$ is marked.}
        \end{figure*}
        %TODO: explain a little bit about advantages and disadvantages of Array vs. list towers. Look at the documentation in the jdk SL.

\textit{\textbf{Search.}} When searching for an element with a key $k$ in the skiplist, the traversal begins from the uppermost level of the sentinel node, where at every step, $k$ is compared to the key of the next node in the current level. Then, if $k > next\rightarrow key$, the traversal continues from the next node and otherwise it continues from the level below. The search completes when a descent is attempted at the bottom level. 
If an element with key $k$ exists, it is the element to the right of the last bottom level node reached in the traversal.  
%If an element with key $k$ is indeed in the skiplist, it is the element to the right of the last bottom level node reached in the traversal.
\begin{algorithm}[htbp]
    \caption{Skiplist search} \label{alg:sl_search}
\begin{algorithmic} [1]
    \Function{Search}{$k$} 
    \State $x\gets header$ \algorithmiccomment{search begins from sentinel node}
    \For {$i \gets \text{skiplist height - 1}$ to $0$} \label{step:for_start}
        \While{$x\rightarrow next[i]\rightarrow key < k$}
            \State $x = x\rightarrow next[i]$
        \EndWhile
    \EndFor \label{step:for_end}
    \State $x = x\rightarrow next[0] $
    \If {$x\rightarrow key == k$} 
        \State \Return $x\rightarrow value$
    \Else
        \State \Return $FAILURE$ 
    \EndIf
\EndFunction
\end{algorithmic}
\end{algorithm}
        
\textbf{\textit{Modifying operations.}}
Inserting or deleting an element with key $k$ begins with a search for that key, recording the last node visited in each level (the node with the largest key strictly smaller than $k$) in a "predecessors" array. The node with the smallest key greater than or equal to $k$ can be recorded in a "successors" array or inferred from the \textit{next} field of predecessor nodes. Then, the skiplist modification is done based on the operation type and whether the key was found: a deletion of a non-existent key would fail, whereas an insert of an existing key is treated by some skiplist implementations as a failure while other skiplist implementations update the  value associated with the key. The two options do not impact the {\sl Foresight} optimization. 

If an element with the key requested for deletion is found, the \textit{next} pointers of its predecessors are rewired to point to the nodes to its right in all the levels that the key appears in. 
If an element with the key requested for insertion is not found, a new tower with a randomly chosen height is inserted into the skiplist by making the $next$ pointers of its predecessors in all of the levels it should appear in point to it. The $next$ pointers of the newly inserted tower are set to point to the nodes 
previously pointed to by its predecessors.
%to which its predecessors previously pointed.
        
\subsection{Concurrent Skiplist Designs}\label{sec::skiplist_concurrent}
%TODO: a short paragraph describing each skiplist we use (sequential, optimistic, NHS, fraser)
%In this paper, we propose a cache-conscious optimization for the skiplist that can be incorporated into existing implementations. We evaluate this optimization on several skiplists, described below. The first is the simple sequential skiplist by Pugh,  described above. The remaining implementations are concurrent, and we explain each of them below.  
In this paper, we propose a cache-conscious optimization that is applicable for a variety of skiplist designs. Besides the sequential skiplist by Pugh, described above,
we implemented and evaluated {\sl Foresight} on three concurrent skiplists: Herlihy et al.'s optimistic lock-based skiplist~\cite{herlihy2006provably,herlihy2007simple}, Fraser's lock-free skiplist~\cite{fraser2004practical}, and Crain et al.'s lock-free No Hot Spot skiplist~\cite{crain2013no}. The following briefly explains how each design works.
%Following is a brief explanation regarding how each of them works.
%

\textit{\textbf{{The Optimistic skiplist.} }}This lock-based skiplist with array-based towers employs fine-grained optimistic locking. Each node has a lock of its own and two flags: \textit{marked} and \texttt{fullyLinked}. Searches are performed without locking.   
Modifying operations first search for the required key, while populating the predecessors and successors arrays. Then, if the modifying operation is a delete operation, the key was found in the skiplist in Node $N$, and $N$ is fully linked (i.e., its insertion completed) and unmarked (not logically deleted), then the operating thread tries to lock the node $N$ and check again that $N$ is not marked. If $N$ was not marked, the thread marks it and continues the operation (otherwise, it releases the lock and returns false). 
The operation proceeds by locking $N$'s predecessors at all of its levels and validating them by checking that they are unmarked and indeed point to $N$ as their successor. 
If this validation succeeds, the thread continues with the modification safely- since it owns the locks, no other thread can disrupt it- and removes the element from the skiplist in a top-down manner. If this validation fails, the predecessors' locks are released and a search is performed again in order to find the correct predecessors, which are then locked and validated in the same manner.

Insert is performed similarly: first a search is performed, populating the predecessors and successors arrays. If a non-marked element with the specified key is found, the operation fails. Otherwise, it continues to lock and validate all the predecessors in the levels equal to or lower than the randomly generated top level of the new node.
If this validation succeeds, the operating thread continues to insert the new element into the skiplist in a bottom-up manner, finally setting the node's \textit{fullyLinked} flag. Otherwise, it releases the locks and repeats this process.
Since validation fails only if another thread modifies a node between the search and locking phases,
%Since validation fails only if another thread changes a node between when it was found in the search and when it was locked prior to validation, 
global progress is guaranteed and deadlock is impossible.

\textbf{\textit{{Fraser's skiplist.}}} This lock-free skiplist with array-based towers essentially maintains a lock-free linked list in each level using the CAS instruction and Harris' pointer-marking scheme~\cite{harris2001pragmatic}. 
%In the original algorithm, skiplist searches aid concurrent operations by physically removing logically deleted nodes (nodes with marked \textit{next} pointers). In its Synchrobench adaptation, an optimization is performed: the default search becomes a "weak" search that does not physically delete nodes and is used for read operations (lookup) as well as (optimistically) for the first search that is performed as part of modifying operations. A "strong" search that deletes physical elements will only be performed as the last step in a delete operation, or if the first ("weak"), search performed in a modifying operation was not sufficient, which may happen due to a concurrent modification by another thread (which might change the \textit{next} pointer of a node in the predecessors array, making the information retrieved during the first search stale, necessitating a new one) or due to one of the predecessors found during the "weak" being logically deleted (hence the need for a "strong" search).
The original algorithm uses searches to physically remove nodes that have been logically deleted (their next pointers were marked). The Synchrobench implementation deploys an optimization using two search types: A "weak" search that does not physically delete nodes, used for read operations and optimistically for the initial search in modification operations; and a "strong" search that physically removes deleted nodes, used only when necessary: as the final step in deletions, or when a weak search becomes insufficient due to concurrent modifications.
The initial weak search may become insufficient in two scenarios: 
(1) concurrent modifications by other threads invalidate the predecessor information, making the search results stale, or (2) predecessors found during the weak search are themselves logically deleted, requiring physical cleanup to be performed by a strong search.

\textbf{\textit{{The No Hot Spot (NHS) skiplist.}}} In this lock-free skiplist variant with linked-list-based towers, the lowest layer (the data layer) consists of a concurrent doubly-linked list that may be updated by all threads. In contrast, the rest of the layers in the skiplist, called the index layers, are only updated by a single designated thread called the \textit{adapting} thread.
Nodes in the data layer contain a key, a value, a \textit{marker} flag (used for deletion) and pointers to the previous and next data nodes.
An index node only contains a pointer to its respective data node and two pointers to the index nodes below it and to its right (or NULL if they do not exist).

Operations are performed by first traversing the index layer to locate an entry point to the data layer, where the operation itself is executed on the data layer list. Importantly, modifying operations only modify the data layer.
The \textit{adapting} thread runs in the background and deterministically decides when to elevate a node's level or to delete the bottom-most level of the index layer in order to maintain the balance of the skiplist.
Thus, the index layer (especially its top-most levels which are traversed frequently) never suffers from high contention.

\section{Foresight}\label{sec::foresight}
%We now present the {\sl Foresight} optimization, beginning with the algorithm and the motivation behind it.
%, discussed in this section. The next section is dedicated to the extension of {\sl Foresight} for concurrent skiplists and its implementation across different skiplist designs. 
Recall that when traversing a skiplist, once a node is reached, the key of the next node must be read to decide whether to advance (move right within the same level) or descend (move down a level). Unfortunately, accessing the next node often results in a cache miss- particularly in large skiplists with many elements or under heavy contention, when concurrent updates are prevalent.
This cache miss can be considered a necessary part of the traversal if it proceeds to the right. However, this is not the case when the traversal instead continues downward. 
We observe that search trees typically do not encounter this problem. With a search tree, only the nodes along the search path need to be accessed. In other words, each node in a search tree stores enough information to determine the next node to access when searching for a given key. %For example, in the case of searching an element in an AVL tree~\cite{AVL} or a B-tree~\cite{B-trees}, 

The {\sl Foresight} optimization brings this desired property to skiplists, making each node contain sufficient data to decide which node should be accessed next in the search path, without necessitating a costly access to another node. 
For a sequential skiplist, this can be done quite simply: at each place where a pointer to a next node is kept, we also keep a $next\_key$ field containing the next node's key, and maintain it accordingly when the next pointed node changes due to insertions or deletions (if there is no next node, we consider the next key to be $\infty$, exactly as in the case without {\sl Foresight}).

A skiplist traversal is a classic example of a pointer-chasing workload: a task in which each step depends on dereferencing a pointer read in the previous step. While extensively researched hardware optimizations such as \textit{data memory-dependent prefetching}~\cite{roth1998dependence, cooksey2002stateless, falsafi2022primer, yu2015imp, ainsworth2016graph, ainsworth2018event, mukkara2018exploiting} can accelerate such workloads, a recent line of work demonstrates that they introduce new security vulnerabilities~\cite{chen2024gofetch, vicarte2021opening, vicarte2022augury}, stressing the need for optimizations at the software, data structure, and algorithmic levels.
        
\begin{figure}[bhtp]
   \centering
    \includegraphics[width=\linewidth]{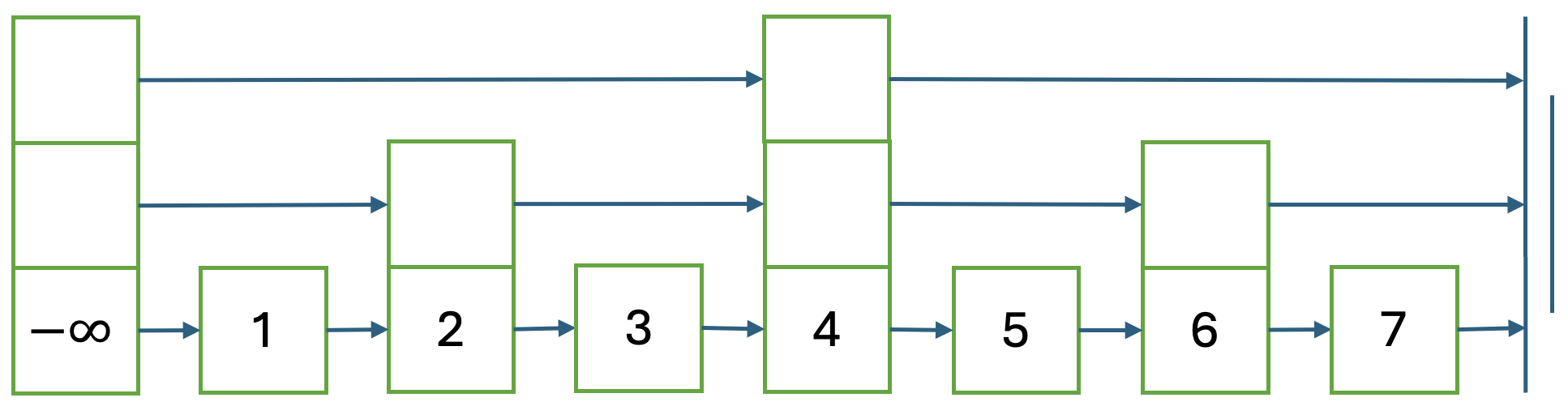}
  \caption{A perfect skiplist illustration.} \label{perfect skiplist figure}
\end{figure}

Our motivation for exploring {\sl Foresight} stems from a rough estimate of the percentage of cache misses it can prevent. Consider skiplists with an expected tower height of $2$ (as in Synchrobench) and for simplicity of the analysis below, assume a perfect skiplist layout (as depicted in Figure~\ref{perfect skiplist figure}). Specifically, there are exactly $2^{\lceil \log n \rceil -i}$ towers of height $i$ ($n$ being the number of elements), with the tallest tower corresponding to the element holding the median key, the second-tallest towers corresponding to the first- and third-quartile keys, and so forth.
Assuming a perfect skiplist simplifies the following motivating analysis.
However, we do not rely solely on this simplified analysis to show the effectiveness of {\sl Foresight}. Instead, we provide an evaluation that includes an empirical measurement of {\sl Foresight}’s impact on cache behavior and execution performance in various skiplist workloads ( presented in Section~\ref{Evaluation}).
        
In a perfect skiplist, a traversal may advance (move right) at most once per level, since the key located two nodes to the right of the starting node at that level also appears in the level above.
Assuming uniformly distributed key accesses, the traversal advances right at each level with probability $0.5$. Thus, the expected number of nodes  visited per level is $1.5$. Without {\sl Foresight}, not only these nodes are accessed: the traversal must also access the node immediately to the right of the last visited node, whose key must be read. Consequently, {\sl Foresight} reduces the expected number of node accesses per level during a traversal from $2.5$ to $1.5$, a $40\%$ reduction.

Another consideration in this approximation is the tower structure. If towers are implemented as linked lists, then the calculations above hold. However, if towers are arrays, the first node visited at each level is not a newly visited node. Moreover, since every other node also appears in a higher level, an access to the third node in the non-optimized algorithm does not constitute a new access, as that node was already accessed one level above before descending. Thus, in each level other than the topmost, only $0.5$ new nodes are expected to be accessed with {\sl Foresight}. Without {\sl Foresight}, with probability $0.5$ the traversal descends immediately after examining the key of the next node to the right, resulting in a single new node access. With probability $0.5$, the traversal continues right to the next node before descending. In this case, the third node is also examined, but it is not new, so the traversal still accounts for exactly one new node access per level. Consequently, the number of new nodes accessed per level decreases from $1$ to $0.5$ with {\sl Foresight}, implying a 50\% reduction due to the optimization.  

When the skiplist contains many elements or experiences %high concurrency 
heavy contention (where nodes are frequently updated by concurrent operations), new node accesses often lead to cache misses. {\sl Foresight} reduces such accesses, thereby reducing misses.
This expectation is corroborated by our experiments in Section $\ref{Evaluation}$.

    \begin{algorithm}[htbp]
            \caption{Sequential skiplist search with Foresight} \label{alg:sl_search}
        \begin{algorithmic} [1]
            \Function{Search}{$k$} 
            \State $x\gets header$
            \For {$i \gets \text{skiplist height - 1}$ to $0$} \label{step:for_start}
                \While{$x\rightarrow next[i].next\_key < k$}
                    \State $x = x\rightarrow next[i].next\_ptr$
                \EndWhile
            \EndFor \label{step:for_end}
            \If {$x\rightarrow  next[0].next\_key == k$} 
                \State \Return $x\rightarrow next[0].next\_ptr\rightarrow value$
            \Else
                \State \Return $FAILURE$ 
            \EndIf
        \EndFunction
        \end{algorithmic}
    \end{algorithm}

\subsection{Concurrency Challenges}
While straightforward in the sequential case, in a multi-core setting, we must ensure that concurrent updates by other threads do not violate the correctness of any operation.
%in a multi-core setting, we must ensure that the correctness of any operation of a thread will not be violated by concurrent updates made by other threads. 
Integrating {\sl Foresight} into a concurrent skiplist introduces risks. 

First and foremost, a difference introduced by {\sl Foresight} is that the \textit{next\_key} field of a node at a certain level is mutable, in contrast to the immutability of each node's \textit{key} field. Thus, \textit{key} must be of a type that can be updated atomically, that is, \textit{key}'s size must be 8 bytes or less in most architectures. We note that many real-world skiplist implementations adhere to this limitation. For example, in LevelDB~\cite{leveldb} and RocksDB~\cite{rocksdb}, skiplist keys are pointers to strings (which allows handling keys of varying sizes comfortably) while in Redis~\cite{redis} skiplist keys (scores) are of type double.

In addition to this limitation, {\sl Foresight} requires each node to contain two fields that must be updated whenever the node to its right at a given level changes: both \textit{next} and \textit{next\_key} rather than just \textit{next} without {\sl Foresight}. Even if these two fields are updated atomically, a problem arises if they are not read atomically: a thread may read $next$ and $next\_{key}$ separately and observe an inconsistent state where $next\_{key} \neq next \rightarrow key$ because both fields were modified between the two reads.
%We first discuss the risks that may result from such inconsistent reads, and then explain how to address this challenge.
Thus, a thread performing a traversal may rely on a foreseen \textit{next\_key} that differs from the actual key of the node to the right (pointed to by \textit{next}). This discrepancy can lead to one of the following two problems.

\textbf{\textit{{Reckless Advance.}}}
During a skiplist traversal, the traversing thread locates the rightmost node at each level whose key is smaller than the search key ($k$). If the foreseen key of the next node to the right is smaller than its actual key, the traversal may incorrectly continue to the right when it should instead descend to the level below (specifically, when $next\_key < k \leq next \rightarrow key$). We refer to this erroneous behavior as \textit{Reckless Advance}.
The invariant of skiplist traversal is that the searched key must always be greater than the keys of the nodes already traversed. Once this invariant is broken, the traversal may miss a key even if it is present in the skiplist. In the case of \textit{Reckless Advance}, a search may fail to find an existing element by “overshooting”, violating correctness. 
%A correct integration of Foresight into a concurrent skiplist must address this issue.
        
\textbf{\textit{{Premature Descent.}}}
In the case symmetric to \textit{Reckless Advance}, an incorrect foreseen value may cause a traversing thread to descend instead of moving right (specifically, when $next\_key \geq k > next \rightarrow key$). We refer to this erroneous behavior as a \textit{Premature Descent}.
If a \textit{Premature Descent} occurs at the bottom level, the search may terminate before reaching a node with $key \geq \textit{searched key}$, causing an existing element to be missed and thereby violating correctness. At higher levels of the skiplist, however, a premature descent does not compromise correctness, since the search will eventually continue rightward until it reaches the node to which the descent should have occurred. In such cases, the impact is limited to a minor efficiency cost, as more nodes will need to be visited in the lower level reached prematurely.
However, in most skiplist implementations, particularly those with array-based towers, when a node is inserted or deleted, the thread performing the operation first traverses the skiplist, recording the last node visited at each level (and sometimes also the successors). These recorded nodes serve as the predecessors needed to determine where a new node should be inserted, or as the predecessors and successors that determine between which neighbors an existing node should be removed, at all relevant levels.
During this traversal, a \textit{Premature Descent} at a level where the node exists (or will exist) may have catastrophic consequences: the new node could be linked into the skiplist at the wrong location, or an incorrect node could be unlinked. 
%This issue too must be addressed by a correct Foresight integration.

\subsection{Solutions}
We propose two techniques to address the aforementioned hazards. The first, \textit{Optimistic Validation}, is general and can be applied to any skiplist design on any platform. The second leverages SIMD instructions and is applicable when the %platform’s vector registers support atomic reads of at least two adjacent words.
platform supports atomic reads and writes of two adjacent memory words.

Incorporating Foresight via either technique introduces only a constant number of additional steps per step of the original algorithm, 
and thus does not weaken the progress guarantee of the underlying skiplist. Specifically, each augmented skiplist preserves the progress guarantee of the original design: 
lock-freedom for Fraser's and NHS's skiplists, and deadlock-freedom for the Optimistic skiplist. If a wait-free skiplist were to be augmented with Foresight
using one of the techniques, it would remain wait-free.
        
\textbf{\textit{{Optimistic Validation.}}} 
%We chose to solve the concurrency hazards that may be caused by Foresight optimistically. That is, we carry out the operations 
%We start with handling \textit{Reckless Advance}. Note that with this error we worry about the case that we have advanced to the next node at the current level (proceeded right) while we should have descended to the lower level because the key of the next node does not match the forecated value and it is too large. As explained above, this happens with low proability due to a rare race. We therefore 
Using this technique, the foreseen keys are optimistically considered to be correct until this assumption may lead to dangerous, correctness-threatening behavior. Then, the real keys are used for validation.
We use this during search to avoid \textit{Reckless Advance} in the following way: after a decision has been made to advance right (Line \ref{Foresight decision} of Algorithm \ref{alg:optimistic validation}), a validation check is performed to ensure that the actual key to the right is indeed smaller than the searched key (Line \ref{validation} of Algorithm \ref{alg:optimistic validation}). If this validation fails, the traversal does not continue to the right.
\begin{algorithm}[htbp]
    \caption{Optimistic Validation Foresight skiplist search} \label{alg:optimistic validation}
\begin{algorithmic} [1]
    \Function{Search}{$k$} 
    \State $x\gets header$
    \For {$i \gets \text{skiplist height - 1}$ to $1$} \label{step:for_start}
        \While{$x\rightarrow next[i].next\_key < k$}\label{Foresight decision}
            \State $next = x\rightarrow next[i].next\_ptr$
            \If{$next\rightarrow key \geq k$}\label{validation}
                \State \textbf{break}
            \EndIf
            \State $x = next$
        \EndWhile
    \EndFor \label{step:for_end}
    \algorithmiccomment{level 0 traversal is done without {\sl Foresight}}
    \State $next = x\rightarrow next[0].next\_ptr$
    \While{$next\rightarrow key < k$}
            \State $x = next$
            \State $next = x\rightarrow next[0].next\_ptr$
    \EndWhile
    \If {$next\rightarrow key == k$} 
        \State \Return $next\rightarrow value$
    \Else
        \State \Return $FAILURE$ 
    \EndIf
\EndFunction
\end{algorithmic}
\end{algorithm}
While this validation introduces some overhead, it still preserves most of {\sl Foresight}'s advantages: 
first, as validation can fail only in the unlikely case that a concurrent modifying thread changes the \textit{next, next\_key} fields of a node traversed by the searching thread in a way that interleaves with its reads, it succeeds in most cases and the validation branch is easy to predict.
Second, the fact that this validation succeeds in most cases means that the decision to advance right based on the foreseen \textit{next\_key} is usually correct. Thus, when the key of the next node is read for validation purposes, the accessed node will actually be visited right after, retaining {\sl Foresight}'s benefit of refraining from accessing non-visited nodes.
        
The same approach of \textit{Optimistic Validation} can be incorporated to cope with incorrect population of the predecessors and successors arrays as a result of \textit{Premature Descent}: instead of preventing \textit{Premature Descent} during search,  the predecessors (and successors) can be validated after a search is performed as part of a modifying operation.    
This approach aligns naturally with the logic already employed by some concurrent skiplists with array-based towers~\cite{herlihy2007simple, fraser2004practical} to validate the predecessors and successors found in a search. We expand on this in Section \ref{Sec: Our Implementations}.
Finally, recall that at the lowest level, a \textit{Premature Descent} can lead to incorrect search results; therefore, we do not apply {\sl Foresight} at this level. At higher levels, we do not explicitly handle \textit{Premature Descent}, since it does not pose a correctness hazard for the search (apart from the case of populating predecessors and successors, addressed above).

\textbf{\textit{{Atomic SIMD Instructions.}}}
%A simple way to ensure that the foreseen key of the next node $x$ of a node $n$ at some level always matches the actual key of $x$, updating the \textit{next} and \textit{next\_key} fields at that level together, and reading them together as well, rendering \textit{Reckless Advance} and \textit{Premature Descent} impossible. 
A simple way to ensure that the foreseen key of the next node $x$ always matches the actual key of $x$ is to update and read the \textit{next} and \textit{next\_key} fields atomically together, rendering \textit{Reckless Advance} and \textit{Premature Descent} impossible.
Writing and reading two adjacent 64-bit fields together is possible on modern Intel CPUs, where aligned 16-byte SIMD loads and stores are guaranteed to be atomic according to Intel's software development manual~\cite{intel64-sdm}. While official documentation from ARM and even AMD does not provide similar atomicity guarantees for 16-byte SIMD operations, empirical evidence~\cite{rigtorp-isatomic} and less formal sources~\cite{llvm-d109827-2021, ibraheem-128bitatomics} suggest that these operations may be atomic on more modern architectures.
%Writing and reading these two fields together is possible, for example, in modern Intel CPUs, where aligned SIMD (Single Instruction, Multiple Data) memory accesses of 16-byte width are guaranteed to occur atomically according to Intel's software development manual~\cite{intel64-sdm} (and where a 16-byte CAS instruction is available). While we could not find similar guarantees for the atomicity of 16-byte-wide SIMD memory accesses in ARM CPUs or even AMD CPUs in official ARM/AMD materials, several informal sources claim that these accesses are indeed atomic.
This approach has limited portability due to its reliance on architecture-specific atomicity guarantees.

\subsection{Integration via Optimistic Validation}\label{Sec: Our Implementations}
We implement {\sl Foresight} for the Fraser~\cite{fraser2004practical}, Optimistic~\cite{herlihy2007simple}, and NHS~\cite{crain2013no} skiplists using \textit{Optimistic Validation}. In all three cases, we prevent \textit{Reckless Advance} by applying \textit{Optimistic Validation} to the search procedure, as shown in Algorithm~\ref{alg:optimistic validation}. 
Recall that we also prevent \textit{Premature Descent} at the bottom level (the data layer) by simply not applying {\sl Foresight} there.
Using {\sl Foresight} without addressing \textit{Premature Descent} further is sufficient to ensure correctness for NHS, since its modifying operations do not rely on predecessors and successors arrays that might otherwise be incorrectly populated during a search due to a \textit{Premature Descent}. 
In the other skiplists, we must still cope with incorrect predecessors and successors arrays that may result from \textit{Premature Descent}.

In the Optimistic skiplist, we do so by adding a new validation criterion (the Optimistic skiplist's validation process is described in Section \ref{sec::skiplist_concurrent}): we now also verify that the successor key at each level is greater than or equal to the searched key. This additional validation step fails only when \textit{Premature Descent} occurs due to concurrent modifications by other threads. Thus, global progress is maintained.
In the original Optimistic skiplist algorithm, if the search in a delete operation finds the node to be deleted at a level that is not its top level, the delete operation fails. This is an optimization, as this node was either marked and partially deleted or not fully linked and the subsequent validation would fail (and the failed delete operation can be linearized right after the delete or right before the insert). With {\sl Foresight}, this is no longer the case since a node can be found in a level which is not its top level due to a \textit{Premature Descent}. Since this optimization is not necessary for the skiplist's correctness (see Reference ~\cite{herlihy2007simple}, Footnote 2), we can and must remove it to preserve correctness with {\sl Foresight}. 

As mentioned above, Fraser's skiplist algorithm includes a strong search, that is, a search procedure that also physically detaches logically deleted nodes it encounters during its traversal. During a strong search, before checking the key of the node to the right, the traversing thread first finds the first node to the right that is not logically deleted, and physically removes all logically deleted nodes in between. As a result, the node to the right must be accessed to check whether it is deleted regardless of checking its key, so {\sl Foresight} does not help in this case. 
Fortunately, as previously discussed, Synchrobench's~\cite{gramoli2015more} implementation of Fraser's skiplist also employs a weak search that does not physically remove nodes, used for lookup operations and for the (optimistic) first search attempt of modifying operations. Therefore, we implement {\sl Foresight} for the weak search only.
Since the weak search does not physically remove logically deleted nodes, and since concurrent modifying operations may corrupt even the predecessors and successors arrays generated by a strong search, Synchrobench's Fraser's skiplist implementation already employs validations of these arrays to decide whether performing a new (strong) search is necessary. We add a new validation to check whether a correctness-threatening \textit{Premature Descent} has occurred during a weak search (ensuring successor keys are not smaller than $k$), initiating a new strong search (without {\sl Foresight}) if it is indeed the case.
Importantly, unlike the NHS and Optimistic skiplists, Fraser's skiplist allows multiple threads to try to change a \textit{next} pointer at the same time using CAS. Thus, we must use a wide (128-bit) CAS to change both \textit{next} and \textit{next\_key} atomically when implementing {\sl Foresight} for Fraser's skiplist, making it less portable as it depends on architectural support for wide CAS.

\subsection{Integration via SIMD Instructions}\label{Sec: SIMD Implementations}
Besides our portable integrations of {\sl Foresight} to concurrent skiplists using \textit{Optimistic Validation}, we also integrated it into all three concurrent skiplists described above (Optimistic, Fraser’s, and NHS) using Intel’s atomic SIMD instructions for synchronization.
Specifically, we make use of the \texttt{MOVDQA} instruction for atomic double-word loads and stores, whose atomicity is explicitly guaranteed in Section 10.1.1 of Intel's Software Development Manual~\cite{intel64-sdm}.
In order to properly use this \texttt{MOVDQA}, we ensure the 16-byte alignment of \textit{next, next\_key} pairs. We note that for Fraser's skiplist, these pairs are also 16-byte aligned when using \textit{Optimistic Validation} to allow usage of wide CAS.

As seen in our evaluation (Section \ref{Evaluation}), this approach can lead to superior performance compared to \textit{Optimistic Validation}. However, SIMD instructions incur higher latency and additional costs (e.g., longer context switches) which may impair performance. 

To the best of our knowledge, this paper is the first to utilize the ability of certain SIMD instructions on specific architectures to atomically load from or store to multiple adjacent memory words. Importantly, this technique is not limited to {\sl Foresight}, and it demonstrates the usefulness of this hardware feature. We hope our work will spark further research into using existing atomic SIMD instructions for efficient synchronization of concurrent algorithms, and incentivize hardware vendors to guarantee the atomicity of more SIMD memory access instructions.

\subsection{Space Overhead}\label{Sec: Space Overhead}
{\sl Foresight} integration into the skiplists in Synchrobench adds space overhead.
With the average tower height being $2$, the average overhead added per skiplist element ranges from $8$ bytes per element incurred when integrating {\sl Foresight} into the Optimistic skiplist using \textit{Optimistic Validation} or into the NHS skiplist using both synchronization methods up 
to $24$ bytes per element when integrating it into the Optimistic skiplist using SIMD or into Fraser's skiplist using both synchronization methods, with integration into the sequential skiplist requiring $16$  bytes per element on average. This is because the average height of a skiplist node is $2$ and in the sequential skiplist we add $8$ bytes for the \textit{next\_key} at each level, while in the Optimistic and NHS skiplists we do not keep this information for the bottom level. Fraser's greater overhead stems from alignment requirements.

The relative space overhead of {\sl Foresight} depends on the underlying skiplist's structure and the size of values indexed by it. When the skiplist implements a set, or a dictionary with small values, {\sl Foresight}'s relative overhead is at its highest. This is exactly the case in the Synchrobench microbenchmarks (discussed in Section \ref{sec::microbenchmarks}), where {\sl Foresight} incurs relative space overheads of $12.5$\% (NHS) to $60$\% (Fraser). Despite these adversarial settings, {\sl Foresight} leads to substantial performance gains. In the DBx1000 macrobenchmarks (discussed in Section \ref{sec::macrobenchmarks}), Fraser's skiplist is used to index table rows of size $100$-$695$B, making {\sl Foresight}'s overhead drop from $60$\% to $3$-$14$\%, demonstrating how {\sl Foresight}'s space overhead decreases in realistic workloads.

 {\sl Foresight}'s overhead  depends on the distribution from which a tower's height is chosen. If the average tower height is lower, {\sl Foresight}'s overhead will be lower as well. {\sl Foresight} introduces a trade-off between performance and space overhead. For example, with \textit{Optimistic Validation}, we do not use {\sl Foresight} on the bottom-most level of the skiplist to avoid correctness-risking premature descent. However, {\sl Foresight} could be disabled in more lower levels in order to reduce its space overhead further. Another design choice that could be made when using a synchronization method where \textit{next} and \textit{next\_key} are always read and changed atomically together (as in our SIMD approach), is to not store a node's key within the node itself, but only next to the pointers to that node within its predecessors (the first node's key can be kept in the sentinel node). To keep our designs as simple and as general as possible, we leave further research into these trade-offs to future work.

\subsection{Safe Memory Reclamation}
In concurrent data structures, nodes may not be reclaimed as soon as they are physically deleted from the data structure, since in-flight operations may still access them. 
To solve this, a Safe Memory Reclamation (SMR) ~\cite{michael2004hazard,fraser2004practical, cohen2015efficient,singh2021nbr} scheme must be deployed, where nodes are first "retired", and they are only reclaimed once they are guaranteed to not be accessed by any operation. 
We use Synchrobench's variant of Epoch Based Reclamation (EBR)~\cite{fraser2004practical} in all of our experiments. 
While {\sl Foresight} is orthogonal to EBR since EBR does not add per-memory-access overhead, and Foresight does not add new nodes to be reclaimed,
there are other popular SMR schemes such as 
Hazard Pointers (HP)~\cite{michael2004hazard} that do introduce per-access overhead, in which case 
{\sl Foresight}'s reduction of accesses may demonstrate a higher positive impact on performance.

\section{Evaluation}\label{Evaluation}
    For our evaluation, we used a dual-socket Intel Xeon Gold 6338 system with 64 physical cores and 128 logical cores total (32 physical cores per socket), with an L1 data cache of 3 MiB total, an L2 cache of 80 MiB total, an L3 cache of 96 MiB total, and 256 GiB of main memory. The operating system and compiler used in our experiments are Ubuntu 24.04.3 and GCC 13.3.0, respectively. Our implementations are in \texttt{C}, matching Synchrobench's \texttt{C} implementations of the sequential, Optimistic, Fraser's and NHS skiplists.

\subsection{Microbenchmarks}\label{sec::microbenchmarks}
    We implemented {\sl Foresight} for four skiplists from the Synchrobench benchmarking suite~\cite{gramoli2015more}: a sequential skiplist~\cite{pugh1990skip}, and the concurrent Optimistic~\cite{herlihy2007simple}, NHS~\cite{crain2013no} and Fraser's~\cite{fraser2004practical} skiplists. We compared the performance of our {\sl Foresight}-augmented implementations against their respective baseline versions.
    Some of the skiplists in Synchrobench implement a \textit{set} interface, while others implement a \textit{dictionary} interface. 
    As {\sl Foresight} affects the underlying skiplist traversal mechanism, and our main goal is to evaluate {\sl Foresight}'s effect on each skiplist's performance rather than compare the different designs to each other, we leave it as is. 
    We modified the Optimistic skiplist implementation: the original Synchrobench algorithm stored the next pointers of a skiplist node in an external array allocated at maximum height, unlike the other implementations of skiplists  with array-based towers (sequential and Fraser's) that use dynamic-sized nodes with internal next pointers. We changed the node struct of the Optimistic skiplist to match this internal pointer layout. This streamlines node structure, achieves significant space savings, and eliminates double dereferencing when accessing successor nodes. This fix alone provides a 1.7× throughput increase (geometric mean across workloads with various skiplist sizes, update ratios, and thread counts). Both our baseline and {\sl Foresight}-augmented Optimistic skiplist evaluations include this fix.
    % \tnote{Optimistic skiplist changes addressed above}
    
    We conducted experiments with varying numbers of participating threads and skiplist sizes, repeating each experiment five times.
    Following common practice in concurrent data structures research, the key range size in each experiment is twice the initial skiplist size, and operations are assigned keys uniformly at random. The number of insertions and the number of deletions are roughly the same in each experiment.
    Following the YCSB guidelines~\cite{cooper2010benchmarking}, we benchmark three different workloads: (1) Read only, (2) $5\%$ insert/remove, $95\%$ read, (3) $50\%$ insert/remove, $50\%$ read. 
    
        \begin{figure*}[htpb]
       \centering
    \includegraphics[width=\linewidth]{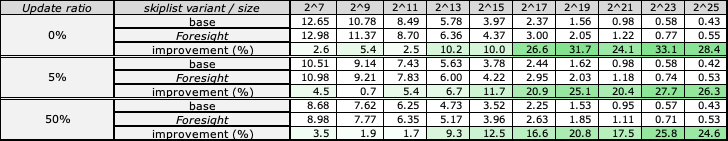}
      \caption{Sequential skiplist microbenchmarks throughput (in Mops) and {\sl Foresight} improvement (as a percentage).} \label{fig::micro_tp_seq}
    \end{figure*}
    
\begin{figure*}[htpb]
       \centering
        \includegraphics[width=\linewidth]{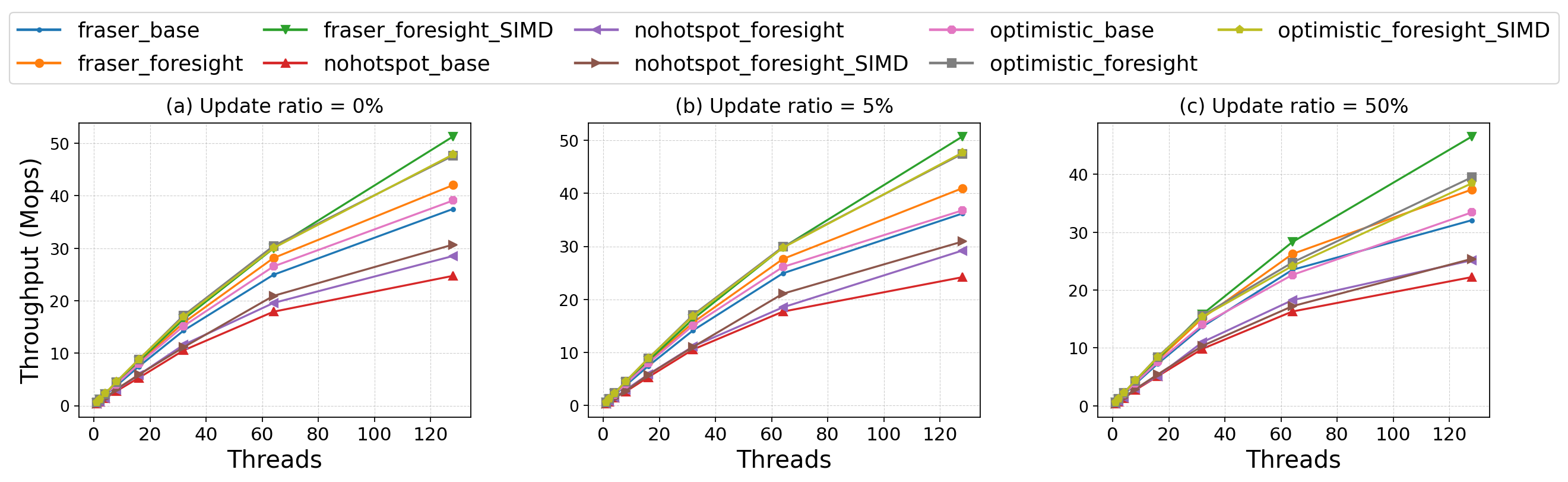}
      \caption{Concurrent skiplists microbenchmarks throughput (in Mops) with data structure size of $2^{25}$ elements, varying number of participating threads.} \label{fig::micro_tp_25}
    \end{figure*}

\begin{figure*}[htpb]
       \centering
        \includegraphics[width=\linewidth]{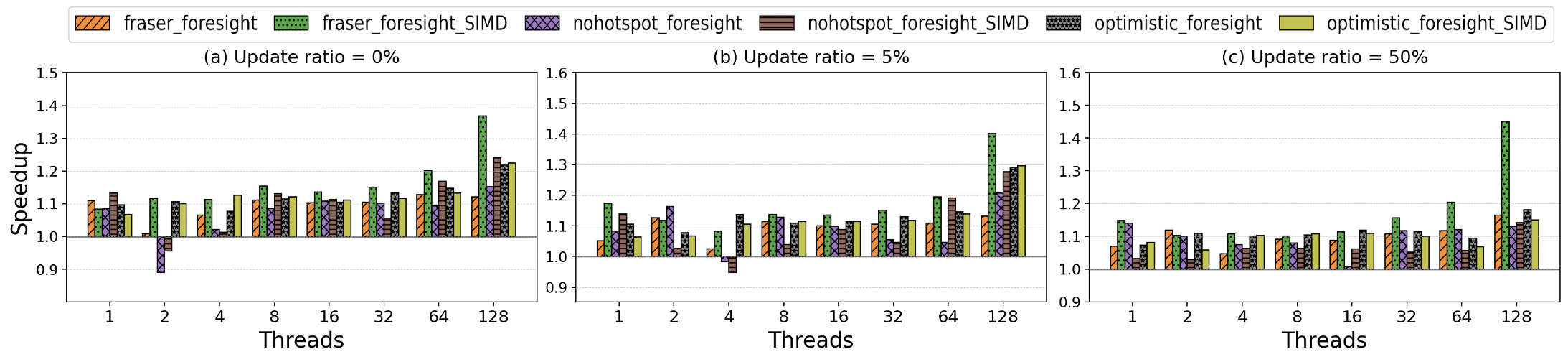}
      \caption{Relative {\sl Foresight} performance improvement when applied to concurrent skiplists, in microbenchmarks with data structure size of $2^{25}$ elements, varying number of participating threads.} \label{fig::micro_su_25}
    \end{figure*}

\begin{figure*}[htpb]
       \centering
        \includegraphics[width=\linewidth]{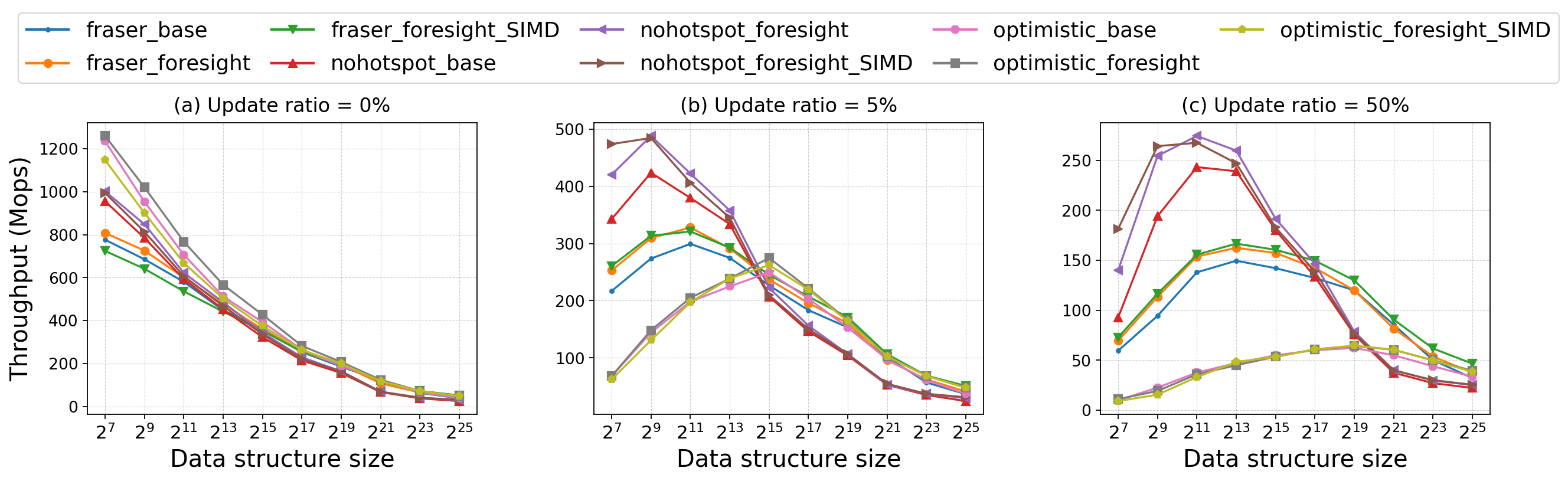}
      \caption{Microbenchmarks throughput (in Mops) with 128 threads, varying data structure sizes.} \label{fig::micro_tp_128}
    \end{figure*}

\begin{figure*}[htpb]
       \centering
        \includegraphics[width=\linewidth]{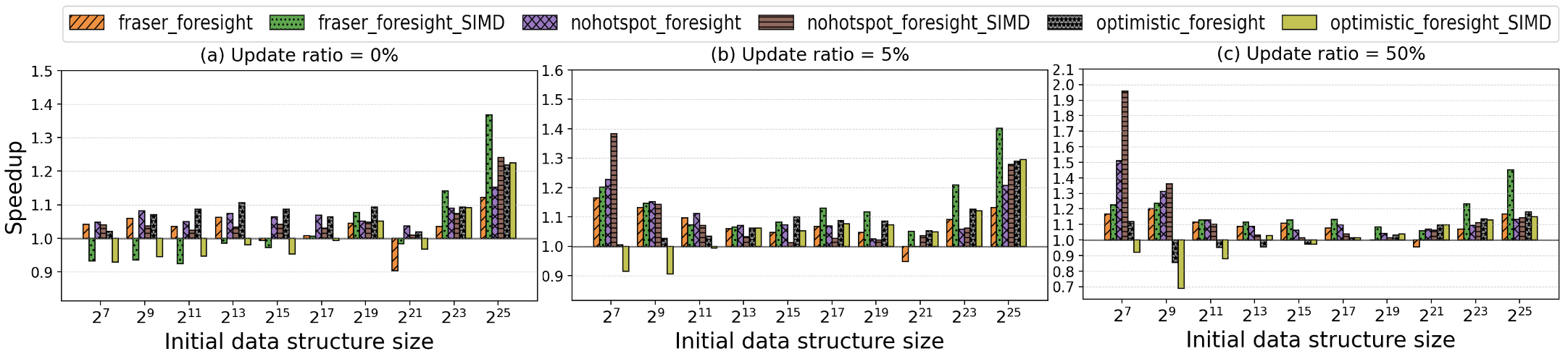}
      \caption{Relative {\sl Foresight} performance improvement in microbenchmarks with 128 threads, varying data structure sizes.} \label{fig::micro_su_128}
    \end{figure*}

    The throughput of the sequential skiplist with and without {\sl Foresight} in the three microbenchmarks with varying data structure sizes is presented in Figure \ref{fig::micro_tp_seq}, together with {\sl Foresight}'s relative speedup. %
    Figures \ref{fig::micro_tp_25} and \ref{fig::micro_su_25} depict the microbenchmark throughput and relative {\sl Foresight} speedup  (respectively) of the different concurrent skiplists, with $2^{25}$ elements and a varying number of participating threads.
    Figures \ref{fig::micro_tp_128} and \ref{fig::micro_su_128} do the same, but with 128 participating threads and varying data structure sizes.
    As shown, {\sl Foresight} scales well with both concurrency and data structure size, providing up to $45$\% speedup for large sizes. When contention is very high (many threads, small data structure size), {\sl Foresight} can significantly improve the throughput of the lock-free NHS and Fraser's skiplists by up to $96$\% (NHS) and $23$\% (Fraser). However, {\sl Foresight}'s impact on the lock-based Optimistic skiplist's performance in high-contention scenarios is not as positive: with Optimistic-Validation-based synchronization it ranges from $-15$\% to $12$\%, and with SIMD-based synchronization it leads to higher losses and no gain. While  {\sl Foresight}'s space overhead might be a plausible explanation for this performance degradation, our cache event monitoring (discussed below) does not reveal more misses incurred by the {\sl Foresight} augmented Optimistic skiplist variants compared to the base variant at any cache level. A plausible explanation for the smaller gain observed in the high-contention case when augmenting the Optimistic skiplist with {\sl Foresight} is its different synchronization mechanism. As can be seen in parts (b) and (c) of Figure \ref{fig::micro_tp_128}, relying on locks severely limits scalability and lowers throughput when contention is high, making cached skiplist nodes less likely to be updated by a concurrent operation (and thus invalidated) before being accessed again in a traversal.

    {\sl Foresight}'s large relative space overhead in the microbenchmarks (described in Section \ref{Sec: Space Overhead}) could explain the performance degradation when the data structure size is $2^{21}$ as seen in Figures \ref{fig::micro_tp_seq} and \ref{fig::micro_su_128}. 
  

To investigate the cause for {\sl Foresight}'s performance impact, we monitor misses across all cache levels with and without {\sl Foresight}.
We used the profiling capabilities of the libpfm~\cite{jarp2008perfmon2} library to measure cache events. 
We ran separate profiling experiments to measure cache events and non-profiling experiments to measure throughput, ensuring that profiling overhead did not affect the reported performance.
% We separated the experiments in which the profiling was used (from which we extract data regarding cache events) from experiments in which it was not used (from which we extract data regarding performance)
% to ensure that the profiling does not affect the measured performance.
%to ensure that the measured performance is not affected by the profiling.
The machine we used supports monitoring of the L1 cache and LLC (L3 cache) only. 
However, the number of L3 references serves as a good approximation for the number of L2 misses.
    
\begin{figure*}[!htpb]
       \centering
        \includegraphics[width=\linewidth]{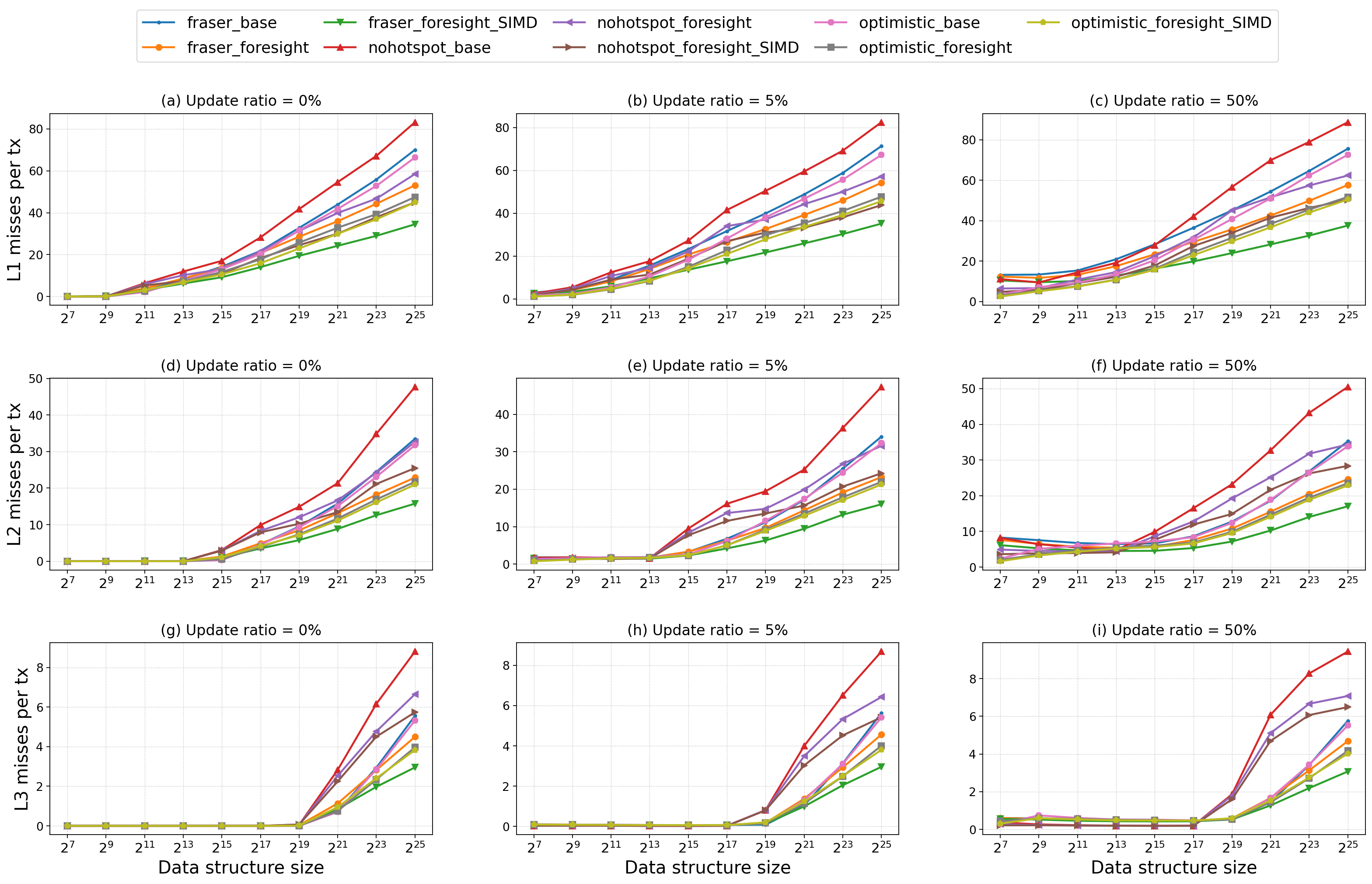}
      \caption{Average cache misses per skiplist operation across workloads (0\%, 5\%, 50\% updates) and cache levels (L1, L2, L3) with 128 participating threads and varying data structure sizes.} \label{fig::micro_cache_128}
    \end{figure*}

    The average number of cache misses per skiplist operation across the different cache levels appears in Figure \ref{fig::micro_cache_128}. The left column  
    presents results for the 0\% update microbenchmark, while the middle 
    and right  
    columns show the 5\% and 50\% update microbenchmarks, respectively. The top, middle, and bottom rows correspond to misses in the L1, L2, and L3 caches, respectively. A considerable cache miss reduction of up to 50\% can be observed in all cache levels when comparing a baseline skiplist design to a {\sl Foresight}-augmented version of the same skiplist. 
    Additional cache events measurements appear in the supplementary material in Appendix \ref{suppM}.

\subsection{Macrobenchmarks}\label{sec::macrobenchmarks}
To evaluate the impact of {\sl Foresight} on a realistic application, we used the DBx1000 in-memory database management system~\cite{yu2014staring}. DBx1000 is widely employed in multi-core in-memory database research~\cite{avni2016persistent, lim2017cicada, yu2014staring, yu2016tictoc, zhang2025rebirth, boeschen2022gacco, chen2024fast}. It implements a relational database with one or more tables and provides multiple mechanisms for concurrency control. Each table consists of multiple rows indexed by an index data structure (a skiplist in our case). For all experiments, we used its most scalable \textit{Two-Phase Locking} (2PL) concurrency-control scheme.
For our experiments, we followed the methodology used in prior work~\cite{arbel2018getting, arbel2018harnessing, nelson2022bundling}, replacing DBx1000’s default index structure with a skiplist. Specifically, we employed a version of Fraser’s skiplist adapted to DBx1000’s index API. We then integrated the {\sl Foresight} optimization to accelerate this skiplist and evaluated its impact on the database’s performance.
We used the following standard workloads.
    
\textbf{\textit{{TPC-C.}}} The Transaction Processing Performance Council’s TPC-C benchmark~\cite{leutenegger1993modeling} aims to simulate a large-scale online business activity. It contains multiple tables of varying sizes (the smallest table contains a row for each warehouse, and the largest contains $100,000$ rows for each warehouse) and is characterized by high contention over row contents. This contention is handled by the database's concurrency control mechanism and is orthogonal to the indexing data structure. DBx1000 supports a representative subset of the TPC-C benchmark, containing only Payment and New Order transactions (which account for $88\%$ of the transactions in TPC-C). We use a number of warehouses equal to the number of threads running the benchmark, as recommended.

\textbf{\textit{{YCSB.}}} The YCSB~\cite{cooper2010benchmarking} workloads use a single table with many rows (either $10M$ or $32M$ in our experiments). Threads access the table rows randomly according to a Zipfian distribution, and either read or update them with a certain probability: $50\%$ read or update in workload A, $95\%$ read and $5\%$ write in workload B and $100\%$ read for workload C (note that these three are the only YCSB workloads properly supported by DBx1000).

\begin{figure*}[!htpb]
       \centering
        \includegraphics[width=\linewidth]{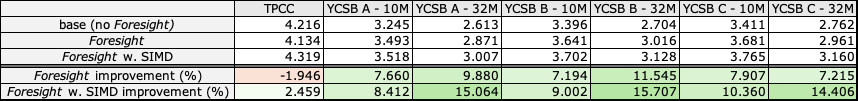}
      \caption{Macrobenchmarks throughput (transactions$/\mu s$)} \label{fig::macro_tp}
    \end{figure*}

We ran the experiments with (1) a (Fraser's) skiplist index without {\sl Foresight} (base), (2) the skiplist index augmented with {\sl Foresight} using \textit{Optimistic Validation} for synchronization and (3) the skiplist index augmented with {\sl Foresight} using SIMD for synchronization.
We ran each experiment with 128 threads until one thread completed $100,000$ transactions (in order to only measure throughput when all threads are active), and repeated each experiment five times. 
The average throughput (in $txns/\mu s$) is reported in the first three rows of Figure \ref{fig::macro_tp}. The last two rows show the relative improvement (as a percentage) of {\sl Foresight} over the base skiplist. As can be seen in the figure, {\sl Foresight} leads to a significant performance improvement of up to $11.5\%$ when using \textit{Optimistic Validation} and up to $15.7\%$ when using SIMD. Unsurprisingly, {\sl Foresight} provides greater throughput gains for benchmarks with larger tables.

    \begin{figure*}[!htpb]
       \centering
        \includegraphics[width=\linewidth]{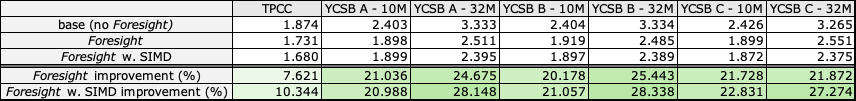}
      \caption{Macrobenchmarks \textit{index time} (seconds per $100,000$ transactions)} \label{fig::macro_it}
    \end{figure*}
    To further investigate the cause for this performance gain, we measure the \textit{index time}, that is, time that threads spend on index operations (i.e., traversing the skiplist). The average \textit{index time} (in seconds per $100,000$ txns) is reported in the first three rows of Figure \ref{fig::macro_it}. The last two rows show the relative reduction in \textit{index time} (as a percentage) over the base skiplist when using {\sl Foresight}. 
    The observed Index time reductions of up to $25.4\%$ (\textit{Optimistic Validation}) and $28.3\%$ (SIMD), corresponding to index-operation speedups of up to $34\%$ and $39\%$, explain the throughput improvements.
\section{Related Work}\label{Related Work}
A common and effective optimization to reduce cache misses in skiplists is packing multiple skiplist elements into skiplist nodes (sometimes referred to as \textit{unrolling}) in order to enhance locality. While originally proposed as a linked list optimization~\cite{braginsky2011locality}, many works implement it for skiplists \cite{platz2019concurrent, sprenger2017cache, xie2016pi, xie2017parallelizing, luo2025bridging}.
Some of these works further capitalize on this unrolling by utilizing SIMD instructions to {\em parallelize} skiplist traversals~\cite{sprenger2017cache, xie2016pi, xie2017parallelizing}. In contrast, our approach leverages the atomicity of SIMD instructions to {\em synchronize} concurrent access.
%While list unrolling list unrolling is an important, well established technique, it's involved nature may complicate its maintenance and integration into existing applications. 
%Unlike list-unrolling, which leads to a distinct, more complicated data structure relying on locks or periodic rebuilding for synchronization, {\sl Foresight} is a simple and general alternative that can be easily applied to existing skiplists and also fits Fraser's skiplist which is lock-free and does not require periodical rebuilding. This may facilitate adoption and maintenance.

List unrolling is a significant reconstruction of the skiplist data structure, yielding a more complex data structure which relies on locks~\cite{luo2025bridging, platz2019concurrent} or periodic rebuilding~\cite{ sprenger2017cache, xie2016pi, xie2017parallelizing} for synchronization, and while it is effective in improving performance, its complexity may hinder wide adoption in practice. For example, all open-source key-value stores cited in this paper use skiplists without unrolling.
In contrast, {\sl Foresight} is a simple, general optimization that can be easily applied to a variety of existing skiplists. In particular, it is compatible with Fraser’s concurrent skiplist,  which requires neither locks nor periodic rebuilding. These properties make the potential adoption of {\sl Foresight} by existing applications easier. 
{\sl Foresight} can potentially be combined with unrolling, although the marginal benefit and {\sl Foresight}'s space overhead may be smaller.
%{\sl Foresight} can potentially be combined with unrolling to further reduce cache misses. While its benefit might be lower (as nodes contain multiple elements, advancing right is likely to occur within the same node), {\sl Foresight}'s space overhead will be lower as well (the nodes are bigger and contain more elements, but {\sl Foresight} will only require storing the smallest key of the node to the right).

The Rotating skiplist~\cite{Rotating} is an extension of the NHS skiplist that addresses the lack of locality of skiplists with linked-list-based towers (where going down in a traversal requires visiting a new node), while maintaining their versatility- namely, the ease of adjusting tower heights. It does so by allocating each skiplist node with the maximal height, and using modulo arithmetic to dynamically control its logical height (replacing static skiplist towers with dynamic, cyclic "wheels"). 
Although the Rotating skiplist demonstrates improvement over NHS in workloads where the skiplist contains up to $64K$ elements, allocating all nodes at maximal height introduces significant space overhead that may hinder performance at larger sizes.
% While demonstrating improvement over the NHS skiplist on
% workloads in which the skiplist contains up to $64K$ elements,
% %, the Rotating skiplist was only evaluated with a relatively small number of elements (up to $64K$). T
% the significant space overhead caused by allocating all the nodes with maximal height %(in contrast to an average node height of $2$ in the NHS skiplist) 
% might hinder performance when the skiplist contains a large number of elements.

NUMASK~\cite{daly2018numask} improves the NHS skiplist for NUMA architectures by maintaining a different skiplist index layer for each NUMA node (while the bottom level containing all the data is shared across all NUMA nodes). {\sl Foresight} can be integrated into NUMASK and the Rotating skiplist as easily as it is integrated into the NHS skiplist.

Moscovici et al.~\cite{moscovici2017gpu} propose a GPU-friendly skiplist algorithm.
Choe et al.~\cite{choe2022hybrids} propose cache-conscious data structures, including a skiplist, designed for Near Memory Processing architectures.

Xing et al.~\cite{xing2025ubiquitous} provide an in-depth survey of the skiplist data structure and its uses, including some of its variants that are  discussed throughout this work. Many previous works discuss cache misses and how to avoid them in other concurrent search data structures, particularly trees \cite{hankins2003effect, rao2000making, mao2012cache, arbel2018getting}. Big Atomics ~\cite{anderson2025big} provides several methods to atomically access several adjacent words together. While they could be used to integrate {\sl Foresight}, they incur higher overhead than the techniques proposed in this paper, and the best performing methods are not lock-free. Big Atomics can facilitate the integration of {\sl Foresight} to a skiplist with keys whose size exceeds a memory word.

Two concurrent works ~\cite{sheng2026architectural, luo2025bridging}, carried out independently in a different setting, used co-location of next-node key with its predecessor's pointer. One of these works uses this optimization for an unrolled skiplist~\cite{luo2025bridging} and preserves thread safety by protecting all operations with locks (including read-only operations). 
This demonstrates that unrolled skiplists can be augmented with {\sl Foresight}. The other work~\cite{sheng2026architectural}, executed in an RDMA environment, uses  transactional memory (TM) to preserve thread safety. They show that co-locating the next key with next pointers leads to noticeable improvements in the RDMA cost model, showing that {\sl Foresight} has merit beyond the multi-core processor setting.

Contrary to these works, this paper derives two integration techniques that can be applied to a wide variety of skiplists with different synchronization methods. Particularly, the techniques presented in this paper are simple and preserve the progress guarantees of the underlying skiplists without relying on complex external features such as transactional memory, locks or background threads (beyond the synchronization features already employed by the underlying skiplists).
This paper also includes a deeper reasoning about the optimization, and a systematic evaluation of it.

\section{Conclusion}
In this paper, we introduced {\sl Foresight}, a cache-friendly optimization for the skiplist data structure. Integrating {\sl Foresight} into a sequential skiplist design is straightforward, but its integration with concurrent skiplists presents additional challenges 
that we analyzed.
We proposed and implemented two approaches to handle these challenges while preserving the progress guarantees of the underlying skiplists: one relying on a combination of optimistic access and validation, and a simpler, more efficient one utilizing
%We proposed and implemented one integration approach, while also noting that a simpler and more efficient alternative is possible with 
SIMD instructions 
on architectures where SIMD instructions can atomically read and write two adjacent memory words.
% where available, since SIMD can atomically read two adjacent words. 
To the best of our knowledge, leveraging the atomicity (in certain modern architectures) of SIMD loads and stores to two adjacent memory words for synchronization purposes has not been previously proposed in the literature. We believe this idea may prove useful in the design of future concurrent data structures.

{\sl Foresight}'s surgical nature makes it widely applicable and easy to integrate. We integrated {\sl Foresight} into a sequential skiplist and three concurrent skiplists: Optimistic, Fraser’s, and the No Hot Spot skiplists. We applied {\sl Foresight} to a skiplist-based index in the DBx1000 in-memory database to evaluate its end-to-end impact. 
While Foresight's integration was lightweight and required only a few dozen lines of code (LoC) changes to augment each skiplist, our evaluation demonstrated throughput improvements of up to 15\% in end-to-end application throughput on DBx1000 and up to 45\% on microbenchmarks.
Our artifacts are available in the github repositories in~\cite{DBx1000_foresight, synchrobench_foresight}. Usage instructions appear in their respective README files.
%indicating that {\sl Foresight} is a practical optimization for skiplist-based indexes in modern in-memory DBMSs.

\section{Acknowledgments}
This work was supported by the Israel Science Foundation Grant No. 3673/25. We thank Gal Shore for his contribution to the implementation of {\sl Foresight} as part of his undergraduate final project.

% use the ACM bibliography style
%\clearpage
\bibliographystyle{ACM-Reference-Format}
\bibliography{sources}
%\clearpage
\appendix
\section{Supplementary Material}\label{suppM}
This appendix contains additional details and measurements of cache events that were excluded from the main paper due to space constraints. The hardware cache events we used to monitor cache references and misses are listed below:
\begin{itemize}
    \item L1 references = L1-dcache-loads + L1-dcache-stores.
    \item L1 misses = L1-dcache-load-misses.
    \item L3 references (L2 misses) = LLC-loads + LLC-stores.
    \item L3 misses = LLC-load-misses + LLC-store-misses.
\end{itemize}

\begin{figure*}[!htb]
    \includegraphics[width=\linewidth]{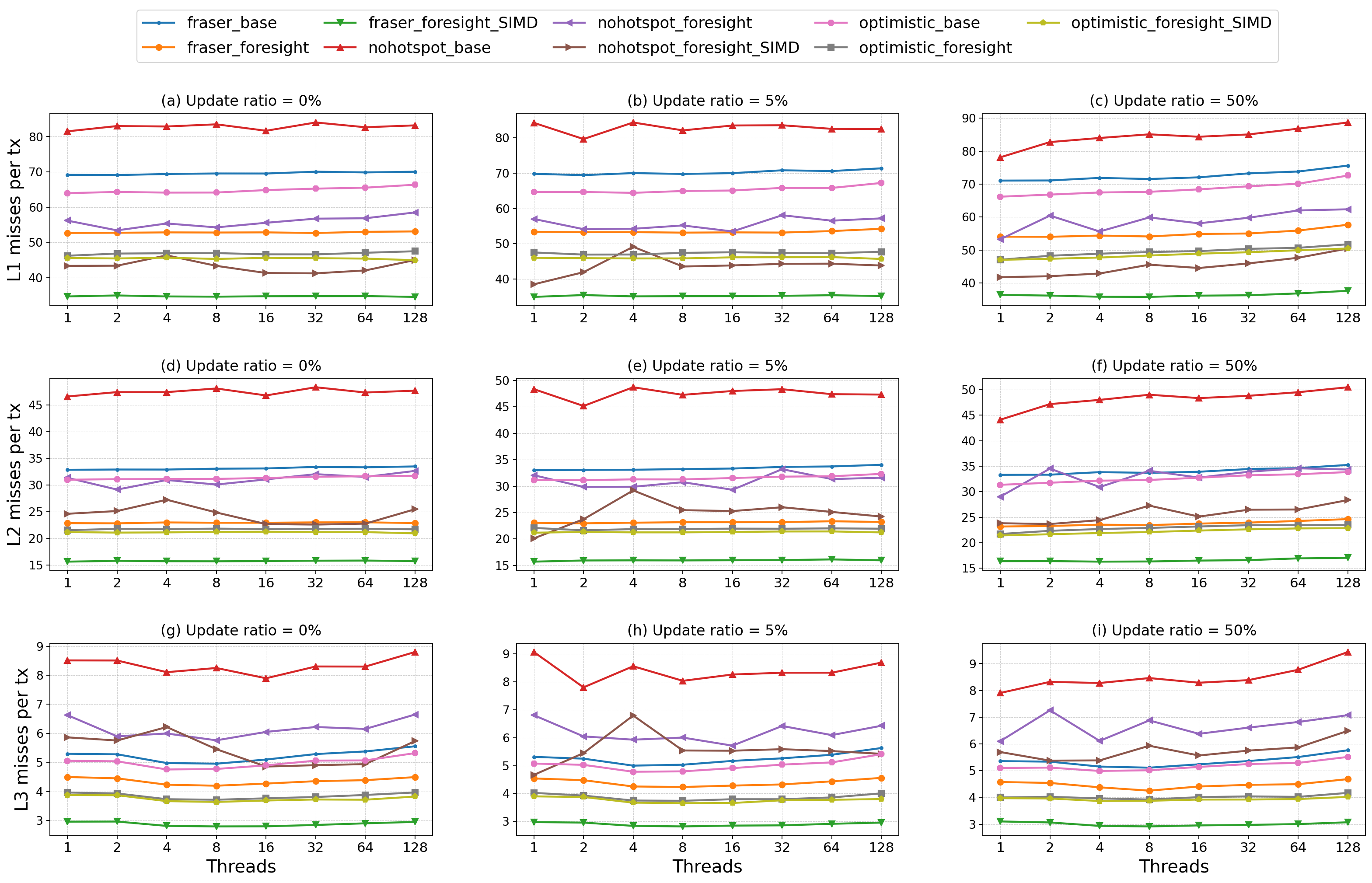}
  \caption{Cache misses per skiplist operation with data structure size of $2^{25}$ elements, varying number of participating threads. The columns correspond to the different microbenchmarks and the rows correspond to the different cache levels.} 
  \label{cache_grid_s25}
\end{figure*}

\begin{figure*}[!htb]
    \includegraphics[width=\linewidth]{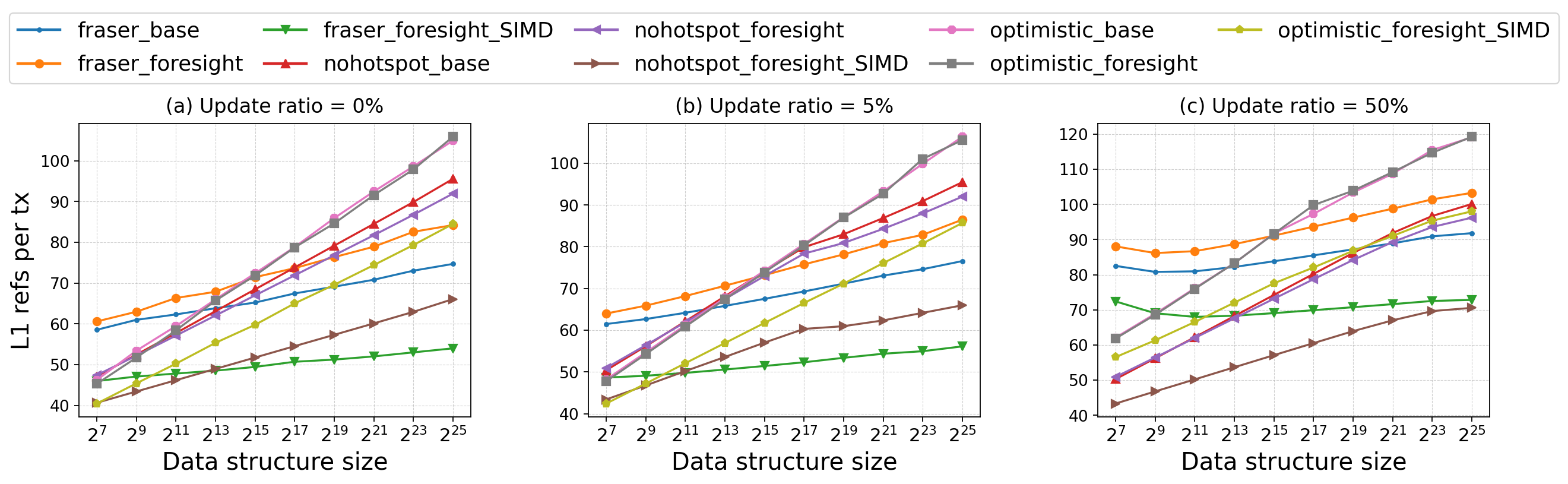}
  \caption{References to the L1 data cache per skiplist operation. 128 threads, varying data structure sizes.} 
  \label{L1_ref_per_op_threads_128}
\end{figure*}

\begin{figure*}[!htb]
    \includegraphics[width=\linewidth]{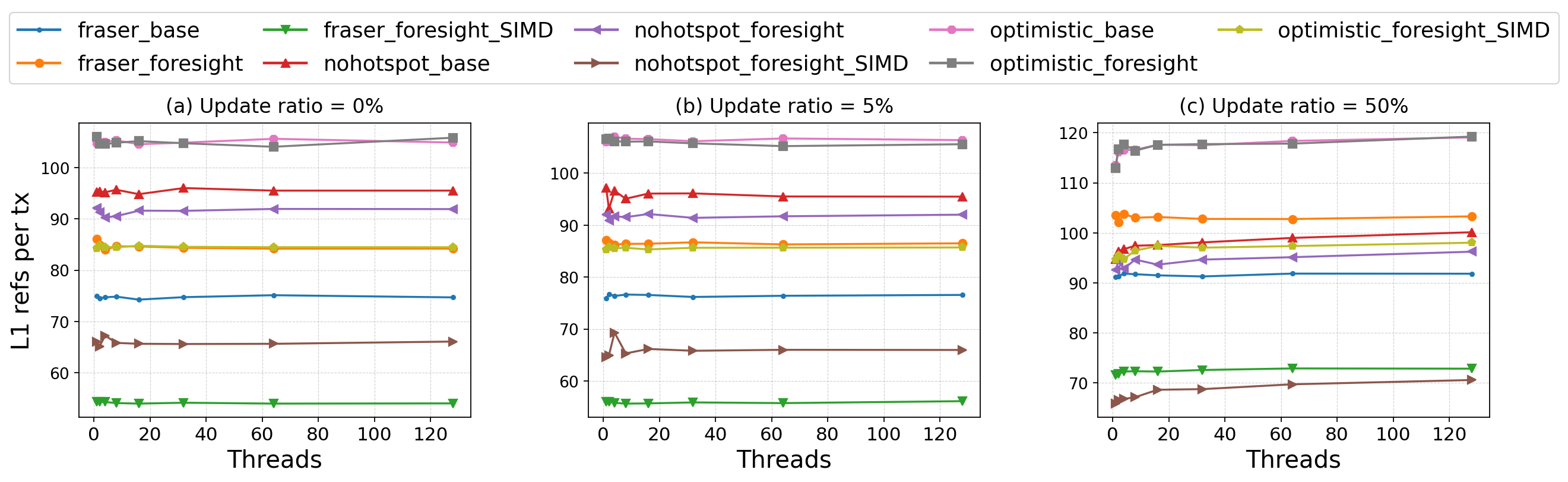}
  \caption{References to the L1 data cache per skiplist operation. $2^{25}$ skiplist elements, varying number of threads.} 
  \label{L1_ref_per_op_size_25}
\end{figure*}

\begin{figure*}[!htb]
    \includegraphics[width=\linewidth]{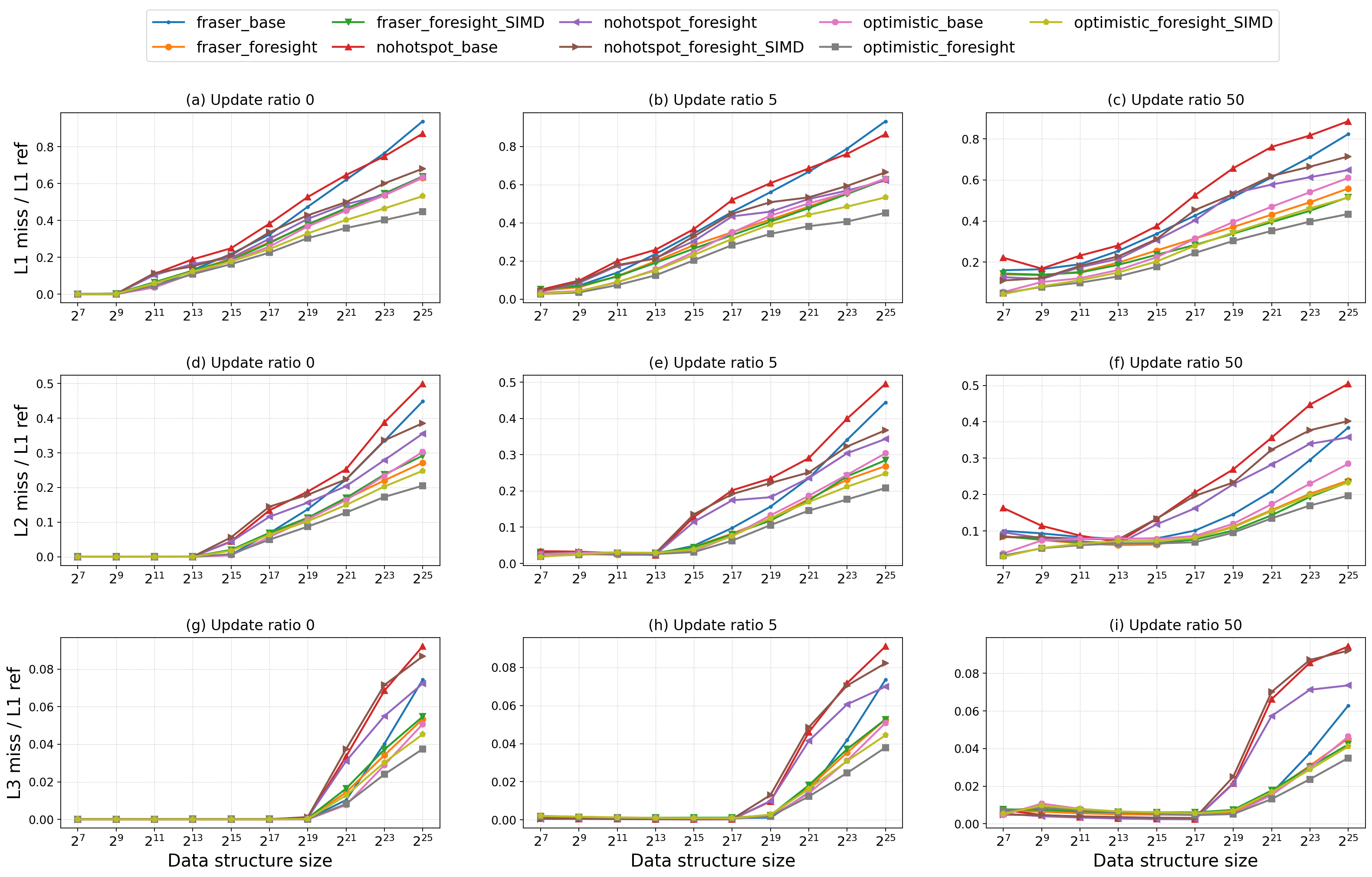}
  \caption{Average number of cache misses per L1 cache reference. 128 threads, varying data structure sizes. The columns correspond to the different microbenchmarks and the rows correspond to the different cache levels.} 
  \label{cache_misses_per_L1_refs_128t}
\end{figure*}

\begin{figure*}[!htb]
    \includegraphics[width=\linewidth]{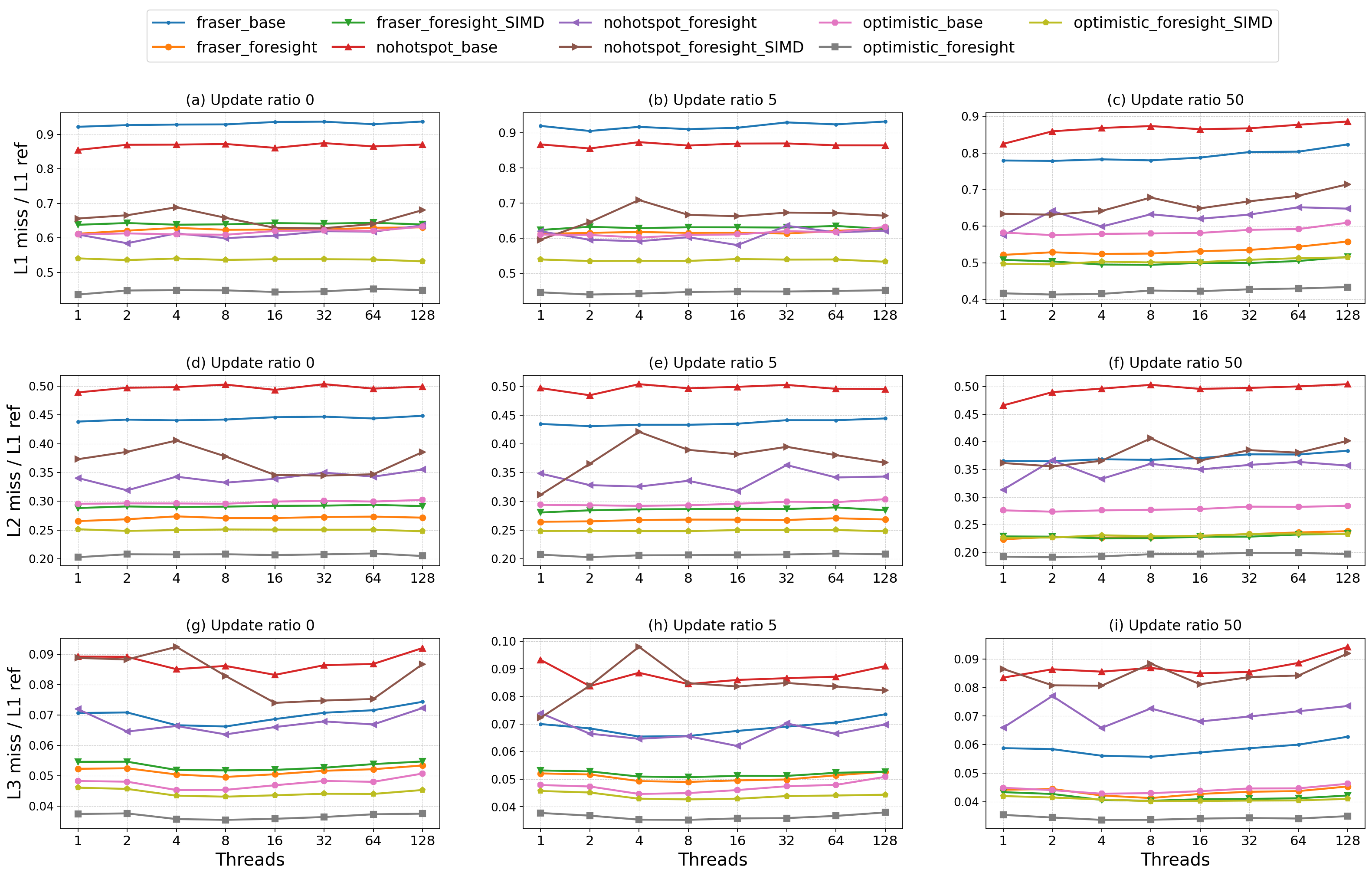}
  \caption{Average number of cache misses per L1 cache reference. $2^{25}$ skiplist elements, varying number of threads. The columns correspond to the different microbenchmarks and the rows correspond to the different cache levels.} 
  \label{cache_misses_per_L1_refs_25s}
\end{figure*}

The additional measurements included in this appendix are: 
\begin{enumerate}
    \item A breakdown of cache misses per operation across cache levels and workloads with $2^{25}$ skiplist elements and a varying number of participating threads (Fig. \ref{cache_grid_s25}).
    \item The amount of references to the L1 data cache per skiplist operation in all experiments (Figures \ref{L1_ref_per_op_threads_128}, \ref{L1_ref_per_op_size_25}).
    \item The percentage of misses in all cache levels out of the total references to the L1 data cache (Fig. \ref{cache_misses_per_L1_refs_128t}, \ref{cache_misses_per_L1_refs_25s}).
\end{enumerate}

%(1) A breakdown of cache misses per operation across all cache levels and all workloads with data structure size of $2^{25}$ elements and a varying number of participating threads (Fig. \ref{cache_grid_s25}), (2) A similar breakdown of cache misses per operation for the sequential skiplist with varying data structure sizes (Fig. \ref{}), (3) The amount of references to the L1 data cache in all the experiments discussed above (Fig. \ref{}), (4) The percentage of misses in all cache levels out of the total references to the L1 data cache (Fig. \ref{}).

Figure \ref{cache_grid_s25} clearly demonstrates the superiority of {\sl Foresight} in terms of reduced cache misses when the skiplist contains a large number of elements.
\end{document}